\documentstyle[twocolumn,prl,aps,floats]{revtex}

\input{psfig.sty}
\begin{document}
\draft
\title{Practical free-space quantum key distribution over 1 km}
\author{W. T. Buttler, R. J. Hughes, P. G. Kwiat, S. K. Lamoreaux,\\
G. G. Luther, G. L. Morgan, J. E. Nordholt, C. G. Peterson, and C. M. 
Simmons}
\address{University of California, Los Alamos National Laboratory \\
Los Alamos, New Mexico 87545}
\date{\today}
\maketitle
%
\noindent{\bf Abstract.} A working free-space quantum key distribution (QKD) system has been 
developed and tested over an outdoor optical path of $\sim1$ km at 
Los Alamos National Laboratory under nighttime conditions. Results 
show that QKD can provide secure real-time key distribution 
between parties who have a need to communicate secretly. Finally, we 
examine the feasibility of surface to satellite QKD.

%
%
Quantum cryptography was introduced in the mid-1980s \cite{ref:BB84} 
as a new method for generating the shared, secret random number 
sequences, known as cryptographic keys, that are used in 
crypto-systems to provide communications security. The appeal of 
quantum cryptography is that its security is based on laws of 
nature, in contrast to existing methods of key distribution that 
derive their security from the perceived intractability of certain 
problems in number theory, or from the physical security of the 
distribution process. 

Since the introduction of quantum cryptography, several groups have 
demonstrated quantum communications {\cite{ref:1_km,ref:TRT}} and quantum 
key distribution \cite{ref:Franson,ref:Marand,ref:ContempPhys,ref:LectureNotes,ref:Gisin,ref:Aero97} 
over multi-kilometer distances of optical fiber. Free-space QKD (over an 
optical path of $\sim 30$ cm) was first introduced in 1991 \cite{ref:32 cm}, 
and recent advances have led to demonstrations of QKD over free-space indoor 
optical paths of 205 m \cite{Buttler}, and outdoor optical paths of 75 m 
\cite{FreeSpace}. These demonstrations increase the utility of QKD by 
extending it to line-of-site laser communications systems. Indeed there are 
certain key distribution problems in this category for which free-space QKD 
would have definite practical advantages (for example, it is impractical to 
send a courier to a satellite). We are developing such QKD, and here we 
report our results of free-space QKD over outdoor optical paths of up to 950 
m under nighttime conditions. 

The success of QKD over free-space optical paths depends on the 
transmission and detection of single-photons against a high 
background through a turbulent medium. Although this problem is 
difficult, a combination of sub-nanosecond timing, narrow filters 
\cite{PhotonByPhoton,DaylightPairs}, spatial filtering \cite{Buttler} 
and adaptive optics \cite{AdaptOptics} can render the transmission 
and detection problems tractable. Furthermore, the essentially 
non-birefringent nature of the atmosphere at optical wavelengths 
allows the faithful transmission of the single-photon polarization 
states used in the free-space QKD protocol.

A QKD procedure starts with the sender, ``Alice,'' generating a 
secret random binary number sequence. For each bit in the sequence, 
Alice prepares and transmits a single photon to the recipient, 
``Bob,'' who measures each arriving photon and attempts to identify 
the bit value Alice has transmitted. Alice's photon state 
preparations and Bob's measurements are chosen from sets of 
non-orthogonal possibilities. For example, using the B92 protocol 
\cite{B92} Alice agrees with Bob (through public discussion) that 
she will transmit a horizontal-polarized photon, $|h\rangle$, for 
each ``0'' in her sequence, and a right-circular-polarized photon, 
$| r \rangle$, for each ``1'' in her sequence. Bob agrees with Alice 
to randomly test the polarization of each arriving photon with 
vertical polarization, $|v\rangle$, to reveal ``1s,'' or 
left-circular polarization, $| \ell \rangle$, to reveal ``0s.'' In this 
scheme Bob will never detect a photon for which he and Alice have 
used a preparation/measurement pair that corresponds to different 
bit values, such as $|h\rangle$ and $|v\rangle$, which happens for 
50\% of the bits in Alice's sequence. However, for the other 50\% 
of Alice's bits the preparation and measurement protocols use 
non-orthogonal bases, such as for $|h\rangle$ and $| \ell \rangle$, 
resulting in a 50\% detection probability for Bob, as shown in 
Table \ref{table:obs_probs}. Thus, by detecting single-photons Bob 
identifies a random 25\% portion of the bits in Alice's random bit 
sequence, assuming a single-photon Fock state with no bit loss in 
transmission or detection. This 25\% efficiency factor, $\eta_Q$, is 
the price that Alice and Bob must pay for secrecy.
\begin{table}[t]
\caption{Observation Probabilities}
\begin{center}
\begin{tabular}{|l|c|c|c|c|} 
Alice's Bit Value & ``0'' & ``0'' & ``1'' & ``1'' \\
Bob Tests With    & ``1'' & ``0'' & ``1'' & ``0'' \\ \hline
Observation Probability & p$=0$ & p$=\frac{1}{2}$ & p$=\frac{1}{2}$ 
& p$=0$ \\ 
\end{tabular}
\end{center}
\label{table:obs_probs}
\end{table} 

Bob and Alice reconcile their common bits through a public 
discussion by revealing the locations, but not the bit values, in 
the sequence where Bob detected photons; Alice retains only those 
detected bits from her initial sequence. The resulting detected bit 
sequences comprise the raw key material from which a pure key is 
distilled using classical error detection techniques. The 
single-photon nature of the transmissions ensures that an 
eavesdropper, ``Eve,'' can neither ``tap'' the key transmissions 
with a beam splitter (BS), owing to the indivisibility of a photon 
\cite{indivis1}, nor copy them, owing to the quantum ``no-cloning'' 
theorem \cite{noclone1}. Furthermore, the non-orthogonal nature of 
the quantum states ensures that if Eve makes her own measurements 
she will be detected through the elevated error rate she causes by 
the irreversible ``collapse of the wavefunction'' 
\cite{Eavesdropping}.
\begin{figure}[t]
\psfig{figure=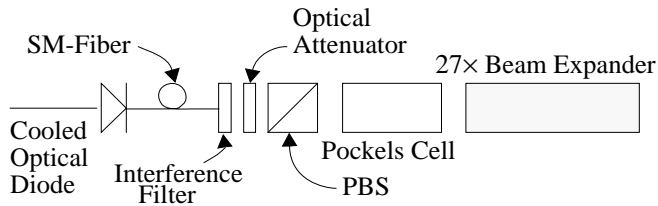,width=3.5 in}
\caption{Free-Space QKD Transmitter (Alice).}
\label{app}
\end{figure} 

The QKD transmitter in our experiment (FIG. 1) consisted of a 
temperature-controlled single-mode (SM) fiber-pigtailed diode laser, 
a fiber to free-space launch system, a 2.5-nm bandwidth 
interference filter (IF), a variable optical attenuator, 
a polarizing beam splitter (PBS), a low-voltage Pockels cell, and a 
$27 \! \times$ beam expander. The diode laser wavelength is temperature 
adjusted to 772 nm, and the laser is configured to emit a short 
pulse of approximately 1-ns length, containing $\sim 10^{5}$ photons.

A computer control system (Alice) starts the QKD protocol by pulsing 
the diode laser at a rate previously agreed upon between herself and 
the receiving computer control system (Bob). Each laser pulse is 
launched into free-space through the IF, and the $\sim 1$ 
ns optical pulse is then attenuated to an average of less than one 
photon per pulse, based on the assumption of a statistical Poisson 
distribution. (The attenuated pulse only approximates a 
``single-photon'' state; we tested out the system with averages 
down to $\lesssim 0.1$ photons per pulse. This corresponds to a 2-photon 
probability of $< 0.5 \%$ and implies that less than 6 of every 100 
detectable pulses will contain 2 or more photons.) 
The photons that are transmitted by the optical attenuator are then 
polarized by the PBS, which transmits an average of less than one 
$|h\rangle$ photon to the Pockels cell. The Pockels cell is randomly 
switched to either pass the light unchanged as 
$|h\rangle$ (zero-wave retardation) or change it to $| r \rangle$ 
(quarter-wave retardation). The random switch setting is determined 
by discriminating the voltage generated by a white noise source.

The QKD receiver (FIG. 2) was comprised of a 8.9 cm Cassegrain 
telescope followed by the receiver optics and detectors. The receiver 
optics consisted of a 50/50 BS that randomly directs collected photons 
onto either of two distinct optical paths. The lower optical path 
contained a polarization controller (a quarter-wave retarder and a 
half-wave retarder) followed by a PBS to test collected photons for 
$|h\rangle$; the upper optical path contained a half-wave retarder 
followed by a PBS to test for $| r \rangle$. One output port along 
each optical path was coupled by multi-mode (MM) fiber to a single-photon 
counting module (SPCM: EG\&G part number: SPCM-AQ 142-FL). [Although the 
receiver did not include IFs, the spatial filtering provided by the MM 
fibers effectively reduced noise caused by the ambient background 
during nighttime operations ($\sim1.1$ kHz) to negligible levels.]
\begin{figure}[t]
\psfig{figure=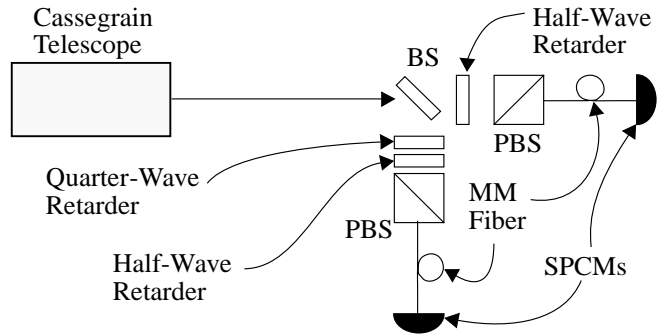,width=3.5 in}
\caption{Free-Space QKD Receiver (Bob).}
\label{a1}
\end{figure}

A single $| r \rangle$ photon traveling along the lower path 
encounters the polarization controller, and is converted to 
$|v\rangle$ and reflected away from the SPCM. Conversely, a single 
$|h\rangle$ photon traveling the same path is converted to 
$| r \rangle$ and transmitted toward or reflected away from the SPCM 
in this path with equal probability. Similarly, a single $|h\rangle$ 
photon traveling the upper path is converted to $|v\rangle$ and 
reflected away from the SPCM in this path, but a single 
$| r \rangle$ photon traveling this path is converted to 
$| \ell \rangle$ and transmitted toward or reflected away from the SPCM 
with equal probability.

The transmitter and receiver optics were operated over 240-, 500-, 
and 950-m outdoor optical paths under nighttime conditions, with the 
transmitter and receiver collocated in order to simplify data acquisition. 
All optical paths were achieved by reflecting the emitted beam from a 
25.4-cm mirror positioned at the half-way point of the transmission 
distance.

The optical coupling efficiency between the transmitter and receiver for 
the 950-m path was $\eta \sim 14\%$, which accounts for losses between 
the transmitter and the MM fibers at the receiver. Bob's detection 
probability, 
\begin{equation}
\label{eq:prob}
P_B \, = e^{-\bar{n}} \, {\displaystyle \sum_{n=1}^{\infty} } 
   \frac{ \bar{n}^n }{n!} \, [1 - y^{n}] = 1 - e^{-\bar{n} \, \eta_B} \, ,
\end{equation}
is the convolution of the Poisson probability distribution of photons in 
Alice's transmitted weak pulse with average photon number $\bar{n}$, and 
the probability that Bob detects at least 1 photon. Here, $y = (1 - \eta_B)$, 
where $\eta_B = \eta \cdot \eta_D \cdot \eta_Q$, and $\eta_D = 65\%$ is 
Bob's detector efficiency. When the transmitter was pulsed at a rate of 20 
kHz with an average of 0.1 photons per pulse for the 950-m path, Eq. 
\ref{eq:prob} gives $\bar{n} \cdot \eta_B = 0.1 \cdot (0.14 \cdot 0.25 
\cdot 0.65) \sim 2.3 \times 10^{-3}$, and hence a bit rate in agreement with 
the experimental result of $\sim 50$ Hz.

The bit error rate (BER, defined as the ratio of the bits received in 
error to the total number of bits received) for the 950-m path was 
$\sim 1.5 \%$ when the system was operating down to the $\lesssim 0.1$ 
photons per pulse level. (A BER of $\sim 0.7 \%$ was observed 
over the 240-m optical path and a BER of $\sim 1.5 \%$ was also observed 
over the 500-m optical path.) A sample of raw key material from the 950-m 
experiment, with errors, is shown in Table \ref{table:key_bits}.
\begin{table}[t]
\caption{A 200-bit sample of Alice's (a) and Bob's (b) raw key 
material generated by free-space QKD over 1 km.}
\begin{center}
\begin{tabular}{|l|c|c|c|c|} 
a & 00000101011101101001000000000001100101010011100010 \\
b & 00000101011101101001000000000001100101010011100010 \\ \hline
a & 01110111011110111000010010001111100000000101101111 \\
b & 01110111011110111000010010001111100000000101101111 \\ \hline
a & 10010010100010000011000001011100001111111111000000 \\
b & 100100101000100000110000010111000011111111{\bf 0}1000000 \\ \hline
a & 10101011011111100111111011110101001101001011101111 \\
b & 10101011011111100{\bf 0}11111011110101001101001011101111 \\ 
\end{tabular}
\end{center}
\label{table:key_bits}
\end{table}

Bit errors caused by the ambient background were minimized to less than 
$\sim1$ every 9 s by the narrow gated coincidence timing windows ($\sim5$ 
ns) and the spatial filtering. Further, because detector dark noise 
($\sim80$ Hz) contributed only about 1 dark count every 125 s, we believe 
that the observed BER was mostly caused by misalignment and imperfections 
in the optical elements (wave-plates and Pockels cell).

This experiment implemented a two-dimensional parity check scheme 
that allowed the generation of error-free key material. A further 
stage of ``privacy amplification'' \cite{PrivAmp} is necessary to 
reduce any partial knowledge gained by an eavesdropper to less than 
1-bit of information, but we have not implemented such a privacy 
amplification protocol at this time. Our free-space QKD system 
does incorporate ``one time pad'' \cite{Vernam} encryption---also 
known as the Vernam Cipher: the only provably secure encryption 
method---and could also support any other symmetric key system.

The original form of the B92 protocol has a weakness to an opaque 
attack by Eve. For example, Eve could measure Alice's photons in 
Bob's basis and only send a dim photon pulse when she identifies a 
bit. However, if Eve retransmits each observed bit as a single-photon 
she will noticeably lower Bob's bit-rate. To compensate for the 
additional attenuation to Bob's bit-rate Eve could send on a dim 
photon pulse of an intensity appropriate to raise Bob's bit-rate to a 
level similar to her own bit-rate with Alice. [In fact, if Eve sends 
a bright classical pulse (a pulse of a large average photon number) 
she guarantees that Bob's bit-rate is close to her own bit-rate with 
Alice.] However, this type of attack would be revealed by our two SPCM 
system through an increase in ``dual-fire'' errors, which occur when 
both SPCMs fire simultaneously. In a perfect system there would be no 
dual-fire errors, regardless of the average photon number per pulse, 
but in an imperfect experimental system, where bit-errors occur, 
dual-fire errors will occur. (We use the dual-fire information to 
estimate the average number of photons per pulse reaching the SPCMs.) 
Our system could also be modified to operate under the BB84 protocol 
\cite{ref:BB84} which also protects against an opaque attack.

Eve could also passively, or translucently, attack the the system using 
a BS and a receiver identical to Bob's (perhaps of even higher efficiency) 
to identify some of the bits for which Alice's weak pulses contains more 
than 1 photon, i.e., Eve receives pulses reflected her way by the BS which 
has reflection probability $R$, whereas Bob receives the transmitted pulses, 
and the BS has transmission probability $T = 1 - R$. Introducing a coupling 
and detection efficiency factor $\eta_E$, for Eve, analogous to Bob's 
$\eta_B$, we find that Eve's photon detection probability is $P_E \, = 
1 - e^{-\bar{n} \, \eta_E \, R}$, whereas Bob's detection probability 
becomes $P_B = 1 - e^{-\bar{n} \, \eta_B \, T}$. (Note: we do not explicitly 
consider any eavesdropping strategy, with or without guessing, in which Eve 
might use more than 2 detectors.

The important quantity is the ratio of the number of bits Eve shares 
with Bob to the number of bits Bob and Alice share. We find that the 
probability that Eve and Bob will both observe a photon on the same 
pulse from Alice is \cite{ref:Aero98,ref:networked}
\begin{equation}
\label{eq:pEB}
P_{B \wedge E} = [1 - e^{-\bar{n} \, \eta_E \, R}] \, 
                 [1 - e^{-\bar{n} \, \eta_B \, T}]\, .
\end{equation}          
To take an extreme case, if Eve's BS has $R = 0.9999$, her efficiency 
is perfect (i.e., $\eta_E = 0.25$), and Alice transmits pulses of 
$\bar{n} = 0.1$, then Eve's knowledge $P_{B \wedge E} / P_B$ of Bob 
and Alice's common key will never be more than 2.5\%. Thus, Alice and 
Bob have an upper bound on the amount of privacy amplification needed 
to protect against a BS attack. Of course, such an attack would cause 
Bob's bit-rate to drop to near zero, and for smaller reflection 
coefficients, $R$, Eve's information on Bob and Alice's key is 
reduced. For example, if Alice transmits pulses with an average of 0.1 
photons per pulse, and $R = T = 0.5$, then for every $250$ key bits 
Alice and Bob acquire, Eve will know only $\sim 3$ bits.

As a final discussion, we consider the feasibility to transmit the 
quantum states required in QKD between a ground station and a satellite 
in a low earth orbit. To that end, we designed our QKD system to operate 
at 772 nm where the atmospheric transmission from surface to space can 
be as high as 80\%, and where single-photon detectors with efficiencies 
as high as 65\% are commercially available. Furthermore, at these optical 
wavelengths depolarizing effects of atmospheric turbulence are negligible, 
as is the amount of Faraday rotation experienced on a surface to satellite 
path.

To detect a single QKD photon it is necessary to know when it will 
arrive. The photon arrival time can be communicated to the receiver by 
using a bright (classical) precursor reference pulse. Received bright 
pulses allow the receiver to set a 1-ns time window within which to look 
for the QKD photon. This short time window reduces background photon 
counts dramatically, and the background can be further reduced by using 
narrow bandwidth filters.

Atmospheric turbulence impacts the rate at which QKD photons would be 
received at a satellite from a ground station transmitter. Assuming 
30-cm diameter optics at both the transmitter and satellite receiver, 
the diffraction-limited spot size would be $\sim 1.2$-m diameter at a 
300-km altitude satellite. However, turbulence induced beam-wander can 
vary from $\sim 2.5\, $--$\, 10$ arc-seconds leading to a photon 
collection efficiency at the satellite of $10^{-3} \,$--$\, 10^{-4}$. 
Thus, with a laser pulse rate of 10 MHz, an average of one photon-per-pulse, 
and atmospheric transmission of $\sim 80\%$, photons would arrive at 
the collection optic at a rate of $800 \,$--$\, 10,000$ Hz. Then, with 
a 65\% detector efficiency, the 25\% intrinsic efficiency of the B92 
protocol, IFs with transmission efficiencies of $\sim 70\%$, and a MM 
fiber collection efficiency of $\sim 40\%$, we find a key generation rate 
of $35 \,$--$\, 450$ Hz is feasible. With an adaptive beam tilt corrector 
the key rate could be increased by about a factor of 100 leading to a key 
rate of $3.5 \,$--$\, 45$ kHz; these rates would double by implementing 
the BB84 protocol.

Errors would arise from background photons collected at the satellite. 
The nighttime earth radiance observed at 300 km altitude at the 
transmission wavelength is $\sim 1$ mW m$^{-2}$ str$^{-1}$ $\mu$m$^{-1}$, 
or $\sim 4 \times 10^{16}$ photons s$^{-1}$ m$^{-2}$ str$^{-1}$ 
$\mu$m$^{-1}$, during a full moon, and drops to $\sim 10^{15}$ photons 
s$^{-1}$ m$^{-2}$ str$^{-1}$ $\mu$m$^{-1}$ during a new moon. Assuming a 
5 arc-seconds receiver field of view, and 1-nm IFs preceding the 
detectors, a background rate of $\sim 800$ Hz (full moon), and $\sim 20$ 
Hz (new moon) would be observed (with a detector dark count rate of $\sim 
50$ Hz, the error rate will be dominated by background photons during 
full moon periods, and by detector noise during a new moon). We infer a 
BER from background photons of $\sim 9 \times 10^{-5} \,$--$\, 10^{-3}$ 
(full moon), and $\sim 2 \times 10^{-6} \,$--$\, 3 \times 10^{-5}$ (new moon). 

During daytime orbits the background radiance would be much larger 
($\sim 10^{22}$ photons s$^{-1}$ m$^{-2}$ str$^{-1}$ $\mu$m$^{-1}$), 
leading to a BER of $\sim 2 \times 10^{-2} \,$--$\, 3 \times 10^{-1}$, 
if an atomic vapor filter \cite{ref:fadof} of $\sim 10^{-3}$ nm bandwidth 
was used instead of the IF. [Note: it would also be possible to place 
the transmitter on the satellite. In this situation, the beam wander is 
similar (2.5--10 arc-seconds), but it is only over the lowest $\sim 2$ km 
of the atmosphere. In this situation, the bit-rate would improve by $\sim 
150$, decreasing the BER by the same amount.]

Because the optical influence of turbulence is dominated by the lowest 
$\sim 2$ km of the atmosphere, our experimental results and this simple 
analysis show that QKD between a ground station and a low-earth orbit 
satellite should be possible on nighttime orbits and possibly even in 
full daylight. During the several minutes that a satellite would be in 
view of the ground station there would be adequate time to generate tens 
of thousands of raw key bits, from which a shorter error-free key stream 
of several thousand bits would be produced after error correction and 
privacy amplification.

This Letter demonstrates practical free-space QKD through a 
turbulent medium under nighttime conditions. We have described a 
system that provides two parties a secure method to secretly 
communicate with a simple system based on the B92 protocol. We 
presented two attacks on this protocol and demonstrated the 
protocol's built in protections against them. This system was 
operated at a variety of average photon number per pulse down to an 
average of $\lesssim 0.1$ photons per pulse. The results were achieved with 
low BERs, and the 240-m experiment demonstrated that BERs of 0.7\% 
or less are achievable with this system. From these results we believe 
that it will be feasible to use free-space QKD for re-keying 
satellites in low-earth orbit from a ground station.

Correspondence and requests for materials to William T. Buttler.
Email: buttler@lanl.gov
R. J. H. extends special thanks to J. G. Rarity for helpful 
discussions regarding free-space quantum cryptography; W. T. B. extends 
appreciation to R. D. Fulton for use of his streak camera and other 
optical diagnostic equipment, to A. G. White for QKD discussions, to G. H. 
Nickel, for turbulence discussions, and to D. F. V. James and P. 
H{\o}yer for translucent BS discussions.
\end{document}